# VDDB: a comprehensive resource and machine learning platform for antiviral drug discovery

Shunming Tao[1, #], Yihao Chen[1, #], Jingxing Wu[1, #], Duancheng Zhao[1], Hanxuan Cai[1], Ling Wang [1, *]

[1] Guangdong Provincial Key Laboratory of Fermentation and Enzyme Engineering, Joint International Research Laboratory of Synthetic Biology and Medicine, Guangdong Provincial Engineering and Technology Research Center of Biopharmaceuticals, School of Biology and Biological Engineering, South China University of Technology, Guangzhou 510006, China.

* **Correspondence**

Ling Wang, South China University of Technology, Guangzhou 510006, China.

Email: lingwang@scut.edu.cn

**Authorship**

# Shunming Tao, Yihao Chen and Jingxing Wu contributed equally to this work.

**Abstract**

   Virus infection is one of the major diseases that seriously threaten human health. To meet the growing demand for mining and sharing data resources related to antiviral drugs and to accelerate the design and discovery of new antiviral drugs, we presented an open-access antiviral drug resource and machine learning platform (VDDB), which, to the best of our knowledge, is the first comprehensive dedicated resource for experimentally verified potential drugs/molecules based on manually curated data. Currently, VDDB highlights 848 clinical vaccines, 199 clinical antibodies, as well as over 710,000 small molecules targeting 39 medically important viruses including SARS-CoV-2. Furthermore, VDDB stores approximately 3 million records of pharmacological data for these collected potential antiviral drugs/molecules, involving 314 cell infection-based phenotypic and 234 target-based genotypic assays. Based on these annotated pharmacological data, VDDB allows users to browse, search and download reliable information about these collects for various viruses of interest. In particular, VDDB also integrates 57 cell infection- and 117 target-based associated high-accuracy machine learning models to support various antivirals identification-related tasks, such as compound activity prediction, virtual screening, drug repositioning and target fishing. VDDB is freely accessible at http://vddb.idruglab.cn.

## INTRODUCTION

Emerging and re-emerging viruses are the pathogens of various terrible infectious diseases of humans and animals, which will continue to threaten public health and result in more premature death than other disease processes and are sustained by global commerce, travel, and disruption of ecosystems.[1,2] Most viruses, such as influenza, hepatitis, dengue, etc., have been known and relatively effectively controlled over the past few decades, but antiviral therapeutics are limited.[3-5] Due to the lack of effective preventive and therapeutic drugs, they may cause pandemics in extreme cases. A typical example is the outbreak of the COVID-19 epidemic on December 31, 2019, which is caused by a new strain of coronavirus termed severe acute respiratory syndrome coronavirus 2 (SARS-CoV-2).[6] SARS-CoV-2 is a highly transmissible and pathogenic coronavirus that has caused a deadly pandemic of acute respiratory disease and continues to ravage the world.[7,8] Many potential therapeutics have been discovered and developed to fight COVID-19, including small molecules that inhibit virus entry or replication within the host, immunomodulators and vaccines.[9] For example, remdesivir, a nucleoside analog that inhibits viral RNA replication by blocking RNA-dependent RNA polymerase, has been approved by the U.S. Food and Drug Administration (FDA) for the treatment of hospitalized patients with COVID-19.[10] Sotrovimab, an approved monoclonal antibody that binds to a highly conserved epitope on the receptor-binding domain (RBD) of the spike protein, can neutralize the new coronavirus and other coronaviruses.[11,12] mRNA-1273 can translate the full-length spike protein and induce T cell-SARS-CoV-2 immune responses in phase I/II/III clinical trials.[13] Although these drugs have shown some benefits in specific subpopulations of patients or certain endpoints, they have not been widely proven to be effective against COVID-19. The most typical example is remdesivir, which has little or no effect on the mortality and length of hospitalization in SARS-CoV-2 patients, according to the solidarity clinical trial data reported by the World Health Organization (WHO).[14]

From the prevention and treatment effects of COVID-19 infection, it can be found that drugs play a key role in the treatment of viral infectious diseases. However, the periodic emergence of viral epidemics and concerns about high costs, long duration, and high mutation rates pose severe challenges for developing novel and effective antiviral drugs.[15,16] The key to drug discovery lies in choosing appropriate therapeutic targets and corresponding drug candidates.[17–19] For example, DNA-dependent RNA polymerase of the pox virus was the first viral enzyme to be discovered in 1967, leading to the discovery of antiviral drugs against pox virus.[20,21] With the in-depth understanding of various viruses, the rapid growth and accumulation of biological, structural, chemical and pharmacological data of these viruses will guide target selection and drug design. Currently, some databases related to antiviral drugs have been reported. In 2008, Xiang et al. presented a web-based resource (VIOLIN) to facilitate the comparison and analysis of vaccines-related research data of human pathogens, including eight viruses (i.e., Ebola virus, influenzae, Hantavirus, Human immunodeficiency virus, Lassa Fever virus, Marburg virus, Vaccinia virus, Venezuelan equine encephalitis virus).[22] In 2014, AVPdb was reported to collect 2683 experimentally-validated antiviral peptides.[23] In 2020, the therapeutic, structural antibody database (Thera-SAbDab) was established to track approximately 460 antibody- and nanobody-related therapeutics recognized by the WHO.[24] Although this database is not explicitly developed for viruses, it contains therapeutic antibody information for some viruses. In 2021, Rajput et al. developed DrugRepV database that stores 3448 unique repurposed drugs and chemicals targeting 23 viruses.[21] Additionally, Raybould et al. reported CoV-AbDab database that contains 1402 anti-coronavirus antibodies and nanobodies known to bind to at least one beta-coronavirus.[25] Clearly, there is still no dedicated resource and platform that provides comprehensive data on antiviral drugs (e.g., vaccines, antibodies, and especially small molecules) for medically important viruses and further data analysis.

To fulfill this gap, we developed VDDB, the first comprehensive dedicated resource to store 715,577 antiviral drugs (i.e., vaccines, antibodies, and small molecules) against 39 medically important viruses. It also contains pharmacological data, mechanism of action, target-related data, various physicochemical and ADMET properties, and literature citations information for the collected entities. In addition, VDDB provides 174 high-precision machine learning (ML)-based predictive models for target and cell-infection assays to support various antiviral drug discovery-related tasks, such as biological activity prediction, virtual screening, compound repositioning, and target fishing.

**SYSTEM CONSTRUCTION AND IMPLEMENTATION**

**Data Collection and Processing.** A schematic of contents, functions and organization is shown in Figure 1. All metadata used in VDDB were manually collected from published scientific literature and free public available databases. First, to obtain information on vaccines and antibodies in clinical trials, a text search was conducted in the US clinical database (ClinicalTrials.gov: https://clinicaltrials.gov/) by using virus or viral as keywords. Only medical drugs were retained from the retrieved information, and then vaccines, antibodies and their associated information such as mechanism of action (MOA) and classification were manually annotated and identified. Second, virus names were used as keywords to search the potential therapeutic targets from TTD (therapeutic target database: http://db.idrblab.net/ttd/, accessed Sep 12, 2021).[26] According to the target Uniprot ID, the corresponding small molecules and their biological activities were retrieved from ChEMBL (https://www.ebi.ac.uk/chembl/, version 29)[27] and PubChem (https://pubchem.ncbi.nlm.nih.gov, accessed Dec 12, 2021)[28] databases. To get more comprehensive drugs/molecules data about the virus, each virus name was also used to capture bioassay information in ChEMBL and PubChem databases. Finally, we also used virus names to retrieve new published papers from PubMed database (https://pubmed.ncbi.nlm.nih.gov/), and then captured drugs/molecules related to these popular viruses. In addition, the potential antiviral targets associated data, including target classification, genes, sequences, functions, structures, pathways, diseases, etc., were retrieved from UniProt[29], TDD, PDB[30], and Reactome[31] databases. Hyperlinks to these external databases were provided for users to inquire about other potentially helpful information.

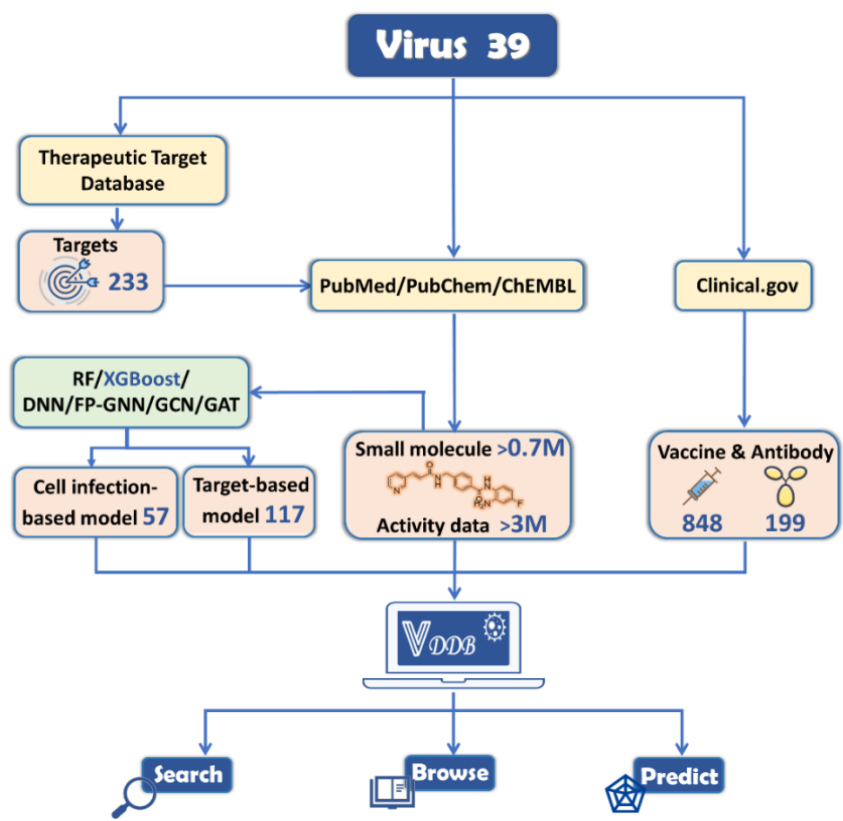

**Figure 1.** Overview of data collection, data processing and database features and contents of VDDB.

**Computational Properties for Small Molecules.** To evaluate the drug-likeness of each molecule stored in VDDB, a series of detailed ligand properties, such as eight physicochemical and twenty-seven ADMET properties were calculated using RDKit (https://www.rdkit.org) and admetSAR tool (http://lmmd.ecust.edu.cn/admetsar1/predict/).[32] Detailed these ligand properties are provided in Table S1.

**Computational Models for Predicting Antiviral Drugs.** To determine whether compounds are active against various viruses and/or their targets (Figure 1), we utilized six ML methods to establish predictive models to detect potential activities. Data selection and ML methods are described below.

**Data Selection.** All quantitative compound-target (assay type: B) and compound-cell (assay type: F) associations were captured from VDDB to build high-quality target- and cell infection-based predictive models for antiviral small molecules. Each target dataset was refined as the following criteria: (i) molecules with biological activity reported as $K_i$, $K_d$, $IC_{50}$, or $EC_{50}$ were retained, while compounds without bioactivity records or whose label could not be unequivocally assigned (e.g., activity < 100 μM or activity > 1 μM) were removed; (ii) the units of bioactivity (i.e., g/mL, M, nM) were converted into the standard unit in μM; (iii) for a molecule with multiple bioactivity values, the final bioactivity value was generated by averaging the available bioactivity records; (iv) according to previous studies,[33–37] compounds with target-based bioactivity values (e.g., $K_i$, $K_d$, $IC_{50}$ or $EC_{50}$) ≤ 10 μM were considered as actives and others were labelled as inactives. Meanwhile, for each cell infection-based antiviral data: (i) only compounds with equivocally assigned bioactivity values (IC50, GI50 or EC50) were kept; (ii) for a compound with multiple bioactivity values, the final bioactivity value was obtained by averaging the available activity values; (iii) compounds with cell-based antiviral activity values (e.g., IC50, GI50 or EC50) ≤ 10 μM were labelled as actives and vice versa.[38–41] All chemical structures in each modelling dataset were standardized using standardizer software (https://github.com/flatkinson/standardiser) with the default parameters, including removing counterions and solvents, neutralizing charge by adding or subtracting atoms, and standardizing structures to a common representation. Once all molecules were standardized, those with molecular weights higher than 800 Da were excluded. Finally, 117 target- and 57 cell-based modelling datasets with at least 30 active compounds were preserved. All datasets used in this study are available at http://vddb.idruglab.cn. Each curated dataset was randomly split into a training set (80%), validation set (10%), and test set (10%). The training set was employed to train the prediction models, the validation set was used for hyperparameter optimization by monitoring predictive power during the training phase, and the test set was performed to evaluate predictive power on unseen data.

**Machine Learning Algorithms and Model Construction.** Two venerable ML methods (Random Forest and Extreme Gradient Boosting) and four advanced deep learning (DL) methods (Deep Neural Networks, Graph Convolutional Network, Graph Attention Network, and FP-GNN) were used to develop predictive models for antivirals. These ML methods are briefly elaborated as follows.

Random Forest (RF) is a representative ensemble learning algorithm to establish a classifier by an ensemble of individual decision trees and to make predictions as final output by vote or by averaging multiple decision trees.[42] RF has a good high tolerance to outliers and noise is not easy to overfit. To obtain the best RF model for each target- or cell-based dataset, five hyperparameters were optimized as follows: n_estimators (10–500), criterion ('gini' and 'entropy'), max_depth (0–15), min_samples_leaf (1–10), and max_features ('log2', 'auto' and 'sqrt').

Extreme Gradient Boosting (XGBoost) is another ensemble learning method under the Gradient Boosting framework and has achieved state-of-the-art (SOTA) ranking results in many ML competitions. It has been widely used in molecular property/activity prediction tasks.[43–45] The following hyperparameters were optimized during the training of XGBoost models: learning_rate (0.01–0.1), gamma (0–0.1), min_child_weight (1–3), max_depth (3–5), n_estimators (50–100), subsample (0.8–1.0), and colsample bytree (0.8–1.0).

Deep neural network (DNN) is a neural network that has several hidden layers. It is made up of several separate neurons, each of which receives information from its related neuron and then activates the aggregated information using a nonlinear activation function.[46] The following key hyperparameters were optimized: dropouts (0.1, 0.2, 0.5), layer_sizes (64, 128, 256, 512) and weight_decay_penalty (0.01, 0.001, 0.0001).

Graph convolutional network (GCN) is a type of neural network that can take in graph-structured data as input and is made up of a graph convolutional layer, a readout layer, a fully connected layer, and an output layer.[47] Duvino et al. developed a convolutional neural network that enables prediction pipeline end-to-end learning.[48] We adopted the GCN method of Duvino et al. and tuned the following hyperparameters: weight decay (0, 10e-8, 10e-6, 10e-4), graph conv layers ([64, 64], [128, 128], [256, 256], learning rate (0.01, 0.001, 0.0001), and dense layer size (64, 128, 256).

Graph attention network (GAT) includes an attention mechanism that automatically learns and calculates the contribution of input data to output data. It solves the inability of the GCN to finish the induction task, that is, dynamic graph problem. The weights of neighboring node features are entirely determined by the node features and are independent of the graph structure.[49] The following hyperparameters were tuned during the training of GAT models: weight_decay (0, 10e-8, 10e-6, 10e-4), learning rate (0.01, 0.001, 0.0001), n_attention_heads (8, 16, 32), and dropouts (0, 0.1, 0.3, 0.5).

FP-GNN is a novel DL architecture for molecular property prediction, which combined and simultaneously learned studied the information from molecular graphs and molecular fingerprints.[50] The evaluation outcomes illustrated that FP-GNN was competitive DL algorithm for molecular property prediction task. To achieve the best FP-GNN model, six hyperparameters are chosen to optimize by using the Hyperopt Pythonpackage[51]: the dropout rate of GAN, the number of multi-head attentions, the hidden size of attentions, the hidden size and dropout rate of FPN, and the ratio of GNN in FP-GNN.

ECFP_4 fingerprint (referred to Morgan fingerprint) was calculated using RDKit software for molecular representation. Morgan fingerprint can record structural environment features of each atom with a diameter of four for each atom and is widely used in QSAR modelling.[33,35,36,39,41,43,52–55] RF models were built using the scikit-learn python package (https://github.com/scikit-learn/scikit-learn, version: 0.24.1), XGBoost models were developed using the XGBoost python package (https://github.com/dmlc/xgboost, version: 1.3.3), FP-GNN models were constructed using the FP-GNN software (https://github.com/idrugLab/FP-GNN, version 1.0), and other graph-based models were established using the DeepChem python package (https://deepchem.io/). The area under the receiver operating characteristic curve (AUC), F1-measure (F1 score), accuracy (ACC), and balanced accuracy (BA) were calculated using the scikit-learn software to evaluate the predictive accuracy of the model.

**Database and Web Interface Implementation.** VDDB is implemented using the open-source LAMP solution stack, which includes MySQL (version 5.6.49) and Apache (version 2.4.46) in the back-end and the Bootstrap (version 3.4.1) framework in the front-end of the web interface. Based on the established predictive models for antiviral molecules, the prediction module of VDDB is developed based on the Flask (version 2.0.3) framework using the Python package. It allows users to input 2D chemical structures in a smiles format or to draw a structure online using JSME (a free molecule editor in JavaScript) [56] for easily and quickly predict the inhibitory activity against various viruses or their targets.

## RESULTS AND DISCUSSION

**Data Contents and Statistics in VDDB.** The content of VDDB is shown in Figure 2. Currently, the collected potential antivirals in VDDB can be divided into three categories, including 848 clinical vaccines (Figure 2A), 199 clinical antibodies (Figure 2B), and 714,530 small molecules (Figure 2C). According to the types of nucleic acid (Figure 2D), these antivirals involved in seven DNA viruses (i.e. Adenovirus (AdV), Cytomegalovirus (CMV), Epstein-Barr virus (EBV), Hepatitis B virus (HBV), Herpes simplex virus (HSV), Human papilloma virus (HPV), Varicella-zoster virus (VZV)), and thirty-two RNA viruses (e.g., SARS-CoV-2, Ebola virus (EBoV), Human immunodeficiency virus (HIV), Dengue virus (DENV), and the Middle east respiratory syndrome coronavirus (MERS-CoV), etc.). The top ten vaccines or antibodies towards their corresponding virus are shown in Figure 3A and 3B. Among different viruses, a total of 236, 180, 129, and 46 vaccines were developed to prevent or treat HIV, Influenza virus (INFV), SARS-CoV-2, and HPV, while the numbers of antibodies against SARS-CoV-2, HIV, HPV, and INFV are 77, 42, 16 and 14, respectively.

In addition, a total of 714,530 unique small-molecule compounds acting on viruses and/or their potential therapeutic targets were integrated in VDDB (Figure 2C). These potential antiviral small molecules are involved in approximately 3 million records of pharmacological data that are measured from 234 experimentally verified targets (Table S2) and 314 cell-based infection screening (Table S3). As shown in Figure 3C, top ten viruses that have been widely studied are SARS-CoV-2, MERS-CoV, SARS-CoV, HIV, HCV, HBV, INFV, EBV, HSV, and Lassa fever virus (LEAVE). Commonly utilized cell to determine inhibitory activity of molecules against various viruses are CHO, HEK293, MT4, Huh-7 and Vero C1008 cells among the top 20 different cell lines (Figure 3D). As illustrated in Figure 3E, 234 potential targets are classified into seven categories based on their roles in the viral infection: entry (17.65%), replication (26.89%), assembly (7.14%), budding (4.62%), immunity (16.39%), inflammation (13.87%), and others (13.45%). Obviously, small molecules targeting viral replication processes and viral enter into host cells represent an attractive strategy to fight against viruses. Furthermore, Figure 3F illustrates that approximately 63.46% of small molecules act on two or more therapeutic targets, indicating that the development of multi-target antiviral drugs is an effective strategy to treat the complex disease of viral infection, which can overcome the shortcomings of single-target drugs that are prone to resistance.

VDDB includes many associated data, such as information related to small molecules (i.e., physicochemical and ADMET properties) and targets (gene, function, sequences, structures, signal transduction pathways etc.). Many data fields in VDDB are hyperlinked to other databases (DrugBank, Reactome, PDB, ChEMBL, PubChem, PubMed, and ClinicalTrials.gov).

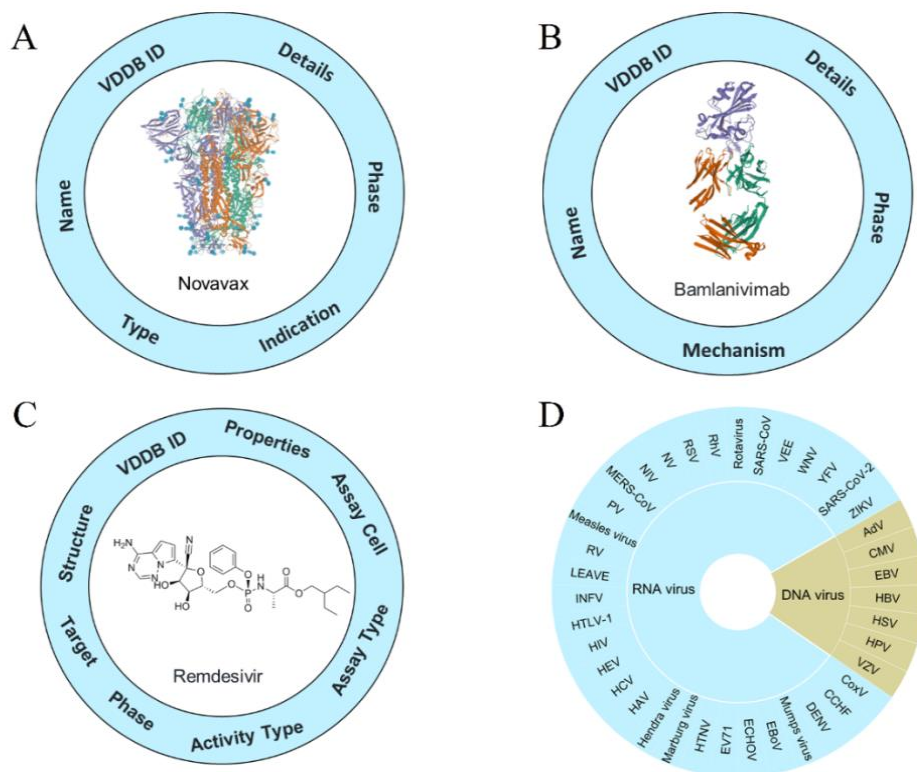

**Figure 2.** Potential antivirals in VDDB. (A) Annotated entry for each vaccine. (B) Annotated entry for each antibody. (C) Annotated entry for each small molecule. (D) Classification of viruses.

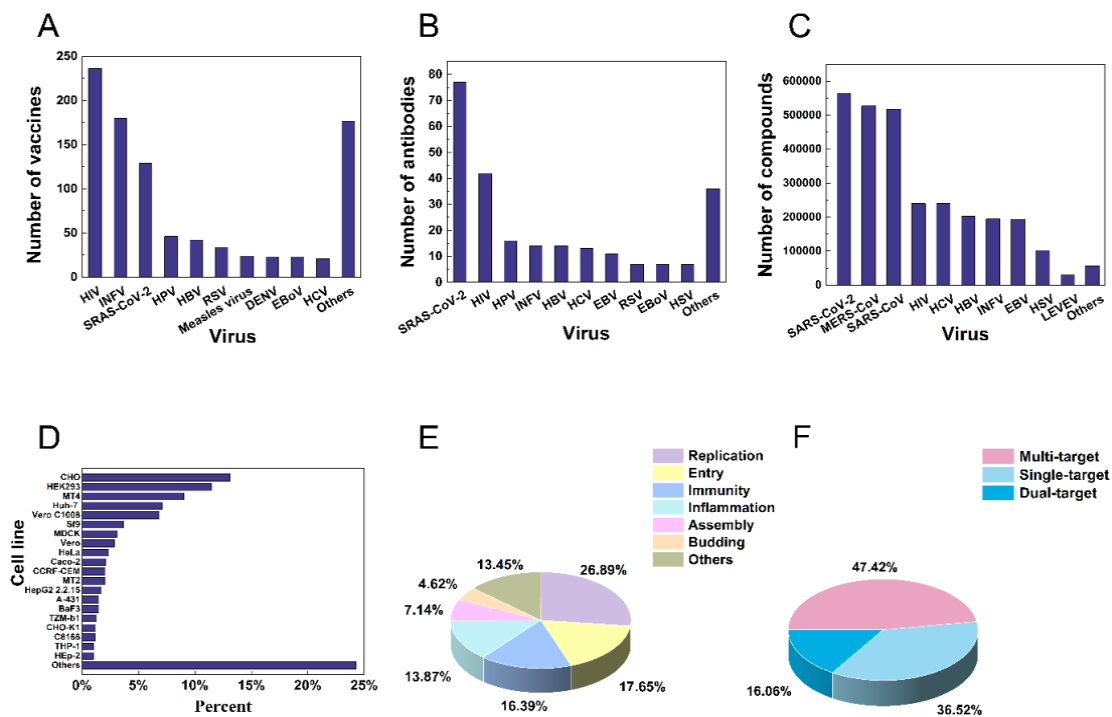

**Figure 3.** Statistics of antiviral resources. (A) The number of vaccines corresponding to the top ten viruses. (B) The number of antibodies corresponding to the top ten viruses. (C) The number of small molecules corresponding to the top ten viruses. (D) Statistics of the top twenty assay cell lines. (E) Classification of targets. (F) Statistics on the proportion of small molecules acting on multiple targets.

**Performance of Predictive Models in VDDB.** 702 predictive models for 117 therapeutic targets against 16 different viruses (Table S4) were built using two CML (RF and XGBoost) and four DL (DNN, GCN, GAT, and FP-GNN) algorithms. Considering the imbalance of actives and inactives in the modelling datasets, F1 score and BA value were adopted to evaluate the accuracy of the model. As shown in Table 1, XGBoost method performed the best overall on these target-based predictive models compared to other ML methods, with the highest average values of F1 (0.862 ± 0.145) and BA (0.620 ± 0.109) for the test set. Meanwhile, XGBoost models achieved the average values of AUC (0.779 ± 0.115) and ACC (0.834 ± 0.116), indicating that the established target-based prediction models can accurately predict the inhibitory activity of molecules against these targets.

In addition, 342 models were constructed to predict antiviral activity of molecules against 57 cell lines involving in 14 different viruses. Detailed performance of these cell infection-based models is listed in Table S4. Compared to other ML methods, XGBoost exhibits the overall best predictive performance on these cell infection-based modelling datasets (Table 1), as it achieves the highest average values of F1 (0.877 ± 0.133), BA (0.752 ± 0.191), AUC (0.856 ± 0.146), and ACC (0.853 ± 0.125) on the test set. Such results imply that the established XGBoost-based predictive models can accurately predict the antiviral activity of compounds on these infected cells. Therefore, the XGBoost-based models were implemented into VDDB platform to predict potential antiviral molecules due to the higher performance of the XGBoost algorithm on both target- and cell infection-based modelling datasets.

**Table 1.** The overall predictive performance of models.

| Models | Test set | | | | |
|---|---|---|---|---|---|
| | Method | AUC[a] | F1[b] | BA[c] | ACC[d] |
| Target-based models | RF | 0.789 ± 0.134 | 0.804 ± 0.300 | 0.575 ± 0.099 | 0.835 ± 0.111 |
| | XGBoost | 0.779 ± 0.115 | 0.862 ± 0.145 | 0.620 ± 0.109 | 0.834 ± 0.116 |
| | DNN | 0.774 ± 0.160 | 0.834 ± 0.251 | 0.614 ± 0.125 | 0.839 ± 0.118 |
| | FP-GNN | 0.748 ± 0.184 | 0.822 ± 0.239 | 0.600 ± 0.130 | 0.785 ± 0.209 |
| | GAT | 0.694 ± 0.185 | 0.772 ± 0.265 | 0.567 ± 0.117 | 0.748 ± 0.188 |
| | GCN | 0.736 ± 0.193 | 0.788 ± 0.259 | 0.599 ± 0.134 | 0.769 ± 0.195 |
| Cell infection-based models | RF | 0.837 ± 0.186 | 0.81 ± 0.238 | 0.609 ± 0.156 | 0.791 ± 0.121 |
| | XGBoost | 0.856 ± 0.146 | 0.877 ± 0.133 | 0.752 ± 0.191 | 0.853 ± 0.125 |
| | DNN | 0.797 ± 0.235 | 0.820 ± 0.217 | 0.608 ± 0.166 | 0.786 ± 0.141 |
| | FP-GNN | 0.687 ± 0.270 | 0.705 ± 0.344 | 0.571 ± 0.124 | 0.711 ± 0.213 |
| | GAT | 0.753 ± 0.212 | 0.655 ± 0.335 | 0.632 ± 0.167 | 0.663 ± 0.232 |
| | GCN | 0.727 ± 0.240 | 0.708 ± 0.283 | 0.594 ± 0.138 | 0.692 ± 0.203 |

[a]AUC: The area under receiver operating characteristic.

[b]F1: F1-measure.

[c]BA: Balanced accuracy.

[d]ACC: Accuracy.

**Usage of the VDDB**

**Search Module.** VDDB can be queried via text mining and chemical similarity searching to acquire relevant data. Two-dimensional (2D) chemical similarity searching is performed based on FP2 fingerprint using RDKit software (http://www.rdkit.org/), and Tanimoto coefficient was used to quantify the similarity score between two molecules. For example, user can enter "hiv" in search bar (Figure 4A), and the related name will be intelligently matched. After clicking submit, information on vaccines, antibodies and small molecules will be displayed. In addition, users can perform similarity search by drawing or pasting a molecule of interest online (i.e., remdesivir), and the database will return compound list results based on a user-selected similarity threshold. More importantly, users can click '+' button in molecular properties function column to display the physicochemical properties and ADMET properties of the small molecule of interest (Figure 4B).

**Browse Module.** In VDDB browse module, users can quickly browse information on clinical vaccines and antibodies, as well as small molecules by clicking on the name of virus or target of interest or mechanism of action of interest in the left function bar (Figure 4C). Taking 3C-like protease (termed as 3CL$^{pro}$ and M$^{pro}$) used for the treatment of COVID-19 as example (Figure 4D), VDDB allows users to browse the basic information about this target (gene, organism, sequence, function, protein structure (available), pathway and indication, etc.) and the corresponding small molecule inhibitors information (structure, mechanism of action, annotated bioactivity, molecular properties, and source, etc.).

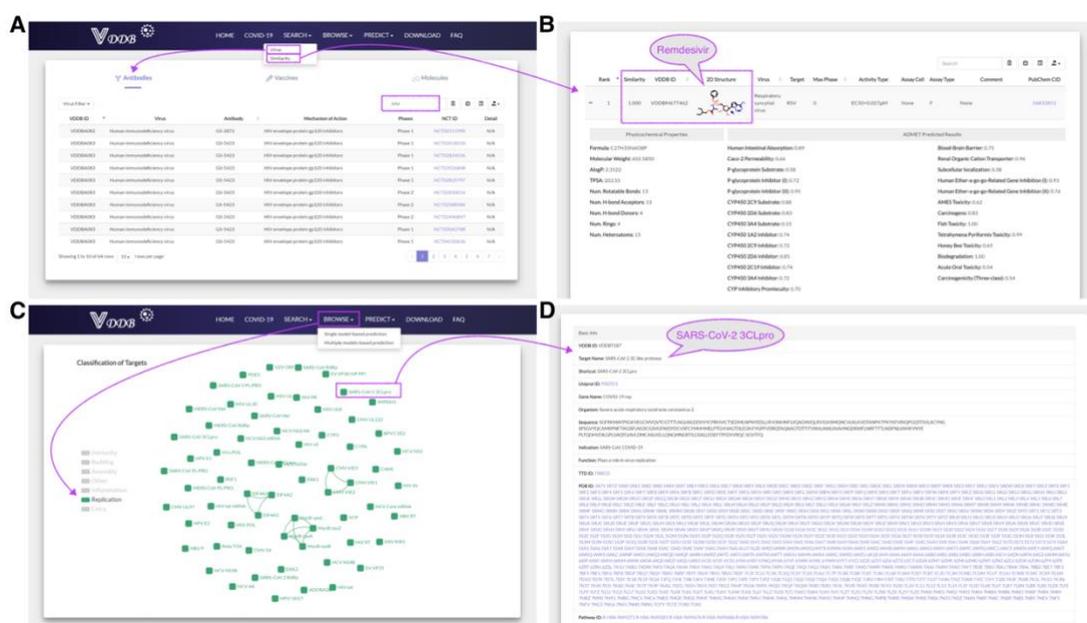

**Figure 4.** Queries and results of the VDDB web-interface. (A) Explaining text mining and search results with example: HIV. (B) Display of chemical similarity search results with example: Remdesivir. (C) Explaining target browse with example: SARS-CoV-2 3CL$^{pro}$. (D) Display of SARS-CoC-2 3CL$^{pro}$ information.

**Predict Module.** VDDB integrates 117 target- and cell infection-based XGBoost models to support virtual screening, drug repositioning and target fishing. Predict module provides two forecasting modes: single model- and multiple models-based prediction. In single model-based prediction (Figure 5A), taking SARS-CoV-2 3CL$^{pro}$ as an example, users first select this target-based model in the left function bar (Figure 5B), and then input molecules of interest (e.g., baicailin, a natural ingredient in Shuanghuanglian which be used to treat COVID-19[57]) by entering or uploading SMILES (limited to 500 molecules ) or by drawing the structure through JSME (Figure 5C). After clicking submit, the predicted score of baicailin against 3CL$^{pro}$ will be generated (Figure 5D, Score = 0.844) that is consistent with the experimental validation results, indicating the prediction accuracy of the target-based model. Meanwhile, Model-related information (Figure 5D), such as model description, modelling dataset and model evaluation metrics can also be displayed online. In addition, each model file (termed Python version software) can be downloaded for users to perform large-scale VS.

In multiple models-based prediction (Figure 5A), users can simultaneously predict the antiviral activity of a given compound against 57 infected cells or 117 targets. Taking the nature product myricetin an example, users can select cell- or target-based models in the left function bar, and then input myricetin by entering or uploading its SMILES or by drawing the structure through JSME (Figure 5C). After clicking submit, the predicted scores of myricetin against all cell infection-based models (Figure 5E) or all target-based models (Figure 5F) will be computed and displayed online. Further analysis of the predicted inhibitory activity of myricetin on virus-infected cells revealed that among the predicted viruses with the highest score, three viruses (HIV, HCV and SARS-CoV-2) could be inhibited by this natural product. For example, myricetin achieves the highest predicted score (0.993) for SARS-CoV-2, indicating that it has inhibitory activity against SARS-CoV-2. Excitingly, such predicted result is confirmed by the latest published research that demonstrates its inhibitory activity against SARS-CoV-2 with an IC$_{50}$ of 0.63 ± 0.01μM.[58] Such results demonstrate the predictive accuracy of the cell infection-based models in real-world drug discovery scenarios.

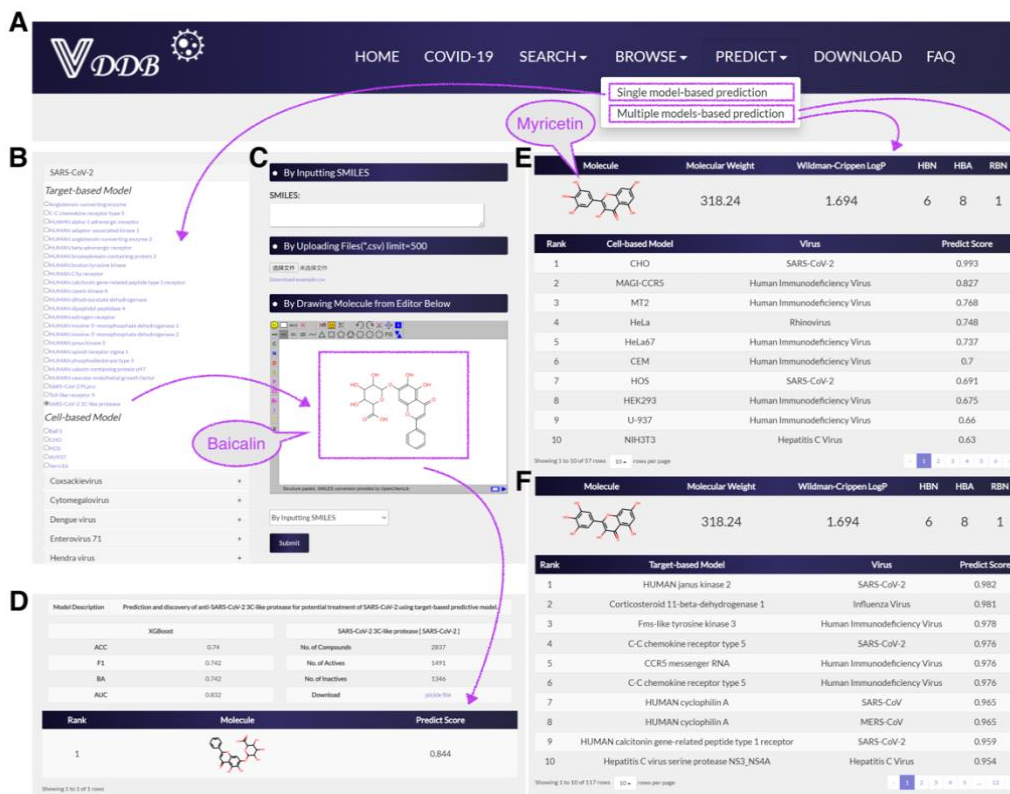

**Figure 5.** Process and cases presentation of VDDB for predicting antiviral activity of small molecules. (A) represents two forecasting methods: single model-based prediction and multiple models-based prediction. (B) represents a single model-based prediction process: the SARS-CoV-2 3C-like protease mode was selected as a case study. (C) represents three ways to input molecular structures. Baicailin was used as an example. (D) represents the detailed predicted results from of the SARS-CoV-2 3C-like protease mode. (E) represents the detailed predicted results for Myricetin from multiple target-based models. (F) represents the detailed predicted results for Myricetin from multiple cell infection-based model.

## CONCLUSION AND PERSPECTIVE

In the face of viral infections or epidemics such as COVID-19, there is an urgent need to collect and share information on existing potential antiviral drugs/molecules to accelerate the rational design and discovery of antiviral drugs. To this end, we developed the first comprehensive web-based platform at this scale dedicated to experimentally-validate antiviral drugs/molecules for 39 medically important viruses including SARS-CoV-2. Extensive antiviral drug-related data and high-accuracy ML-based models in VDDB can serve as a reference for the research community and support various antivirals design and identification tasks. With the continued influx of new potential antivirals, we will continue to add new data as well as new predictive models, periodically update the VDDB, and enhance the usability of the online interface, allowing researchers to rapidly query, browse, acquire data on antivirals, as well as to perform various antivirals design and identification tasks online. We expect that VDDB can be served as a valuable resource and powerful tool for the rational design and discovery of potential drugs to fight viral infectious diseases.